\documentclass{elsart}

\usepackage{graphics}
\usepackage{graphicx}
\usepackage{epsfig}

\def\lta{~\raise.4ex\hbox{$<$}\llap{\lower.6ex\hbox{$\sim$}}~}
\def\gta{~\raise.4ex\hbox{$>$}\llap{\lower.6ex\hbox{$\sim$}}~}

\begin{document}

\begin{frontmatter}



\title{Evolutionary Markovian Strategies in $2 \times 2$ Spatial Games}


\author[label1]{H. Fort}
%\author[label1]{} 
\author[label2]{and E. Sicardi}

\address[label1]{Instituto de F\'{\i}sica, Facultad de Ciencias, Universidad
de la Rep\'ublica, Igu\'a 4225, 11400 Montevideo, Uruguay}
\address[label2]{ Instituto de F\'{\i}sica, Facultad de Ingenier\'{\i}a,
Universidad de la
Rep\'ublica, Julio Herrera y Reissig 565, 11300 Montevideo, Uruguay. }

\begin{abstract}
Evolutionary spatial $2\times2$ games between heterogeneous agents are analyzed 
using different variants of cellular automata (CA). 
Agents play repeatedly against their nearest neighbors $2\times2$ games 
specified by a rescaled payoff matrix with two parameteres.
Each agent is governed by a binary Markovian strategy
(BMS) specified by 4 conditional probabilities [$p_R$, $p_S$, $p_T$, 
$p_P$] that take values 0 or 1. The initial configuration consists in a 
random assignment 
of "strategists" among the $2^4=$ 16 possible BMS. The system then evolves
within strategy space according to the simple standard rule: each agent
copies the strategy of the neighbor who got the highest payoff. Besides on 
the payoff matrix, the dominant
strategy -and the degree of cooperation- depend on i) the type of the
neighborhood (von Neumann or Moore); ii) the way the cooperation state is
actualized (deterministically or stochastichally); and 
iii) the amount of noise measured by a parameter $\epsilon$.
However a robust winner strategy is [1,0,1,1]. 
\end{abstract}

\end{frontmatter}

% main text


\section{Introduction}

$2\times2$ non cooperative games consist in two players 
each confronting two choices: cooperate (C) or defect (D) and each 
makes its choice withoutknowing what the other will do.
The four possible outcomes for the interaction of both agents 
are: 1) they can both cooperate (C,C) 2) both defect (D,D), 3) 
one of them cooperate and the other defect (C,D) or (D,C).
Depending on the case 1)-3), the agent gets respectively
: the "reward" $R$, the "punishment" $P$ or the "sucker's
payoff" $S$ the agent who plays C and the "temptation to defect" $T$ 
the agent who plays D.
One can assign a {\em payoff matrix} M given by
 
\begin{center}
 
${\mbox M}=\left(\matrix{(R,R)&(S,T)\cr (T,S)&(P,P) \cr}\right),$
 
\end{center}
which summarizes the payoffs for {\it row} actions when confronting
with {\it column} actions.

The paradigmatic non zero sum game is the Prisoner's Dilemma (PD).
For the PD the four payoffs obey the
relations: $T>R>P>S$ and $2R>S+T$ \footnote{The last condition is
required in order that the average utilities for each agent of a cooperative
pair ($R$) are greater than the average utilities for a pair exploitative-  
exploiter (($R+S$)/2).}.
Clearly in the case of the PD game it always pays more to defect independently 
of what your opponent does: if it plays D you can got either $P$ (playing D) 
or $S$ (playing C) and if it plays C you can got either $T$ (playing D) or    
$R$ (playing C). 
Hence defection D yields is the {\it dominant strategy} for rational agents.
The dilemma is that if both defect, both do worse than if
both had cooperated: both players get $P$ which is smaller than $R$.
A possible way out for this dilemma is to play the game repeatedly.  
In this iterated Prisoner's Dilemma (IPD), players meet several 
times and provided they remember the result of previous encounters, more
complicated strategies than just the unconditional C or D are possible.
Some of this {\it conditional} strategies outperform the dominant one-shot 
strategy, "always D", and lead to some non-null degree of cooperation.

The problem of cooperation is often approached from a Darwinian evolutionary 
perspective: diverse strategies are let to compete and the most successful 
propagate displacing the others.
The {\it evolutionary game theory}, originated as an application of
the mathematical theory of games to biological issues \cite{msp73},
\cite{ms82}, later spread to economics and social sciences \cite{axel84}.

The evolution of cooperation in IPD simulations may be understood in terms
of different mechanisms based on different factors.
Among the possible solutions a very popular one regards reciprocity 
as the crucial property for a winner strategy.
This was the moral of the strategic tournaments organized by in the early 
eighties by Axelrod \cite{ah81},\cite{axel84}. 
He requested submissions from several specialists in game theory 
from various disciplines. He played first the strategies against each other 
in a round robin tournament, and averaged their scores. 
The champion strategy was {\it Tit for Tat} (TFT): cooperate on 
the first move, and then cooperate or defect exactly as your opponent did on
the preceding encounter.  
Then he evaluated these strategies by using genetic algorithms
that mimic biological evolution.  That is, the starting point is a population
of strategies, with one representative of each 'species', or competitor. If a
strategy performed well, in the next generation it would be represented more
than once, and if a strategy did poorly, it would die off. Again, 
TFT dominated in most of these "ecological" tournaments. 
Axelrod identified as key features for the success of TFT, besides nicety 
(it began playing C in the first move and never is the first to defect on an 
opponent), two facets of reciprocity, namely: a) it retaliates, meaning that 
it did not ignore defection but responded in kind, and b) forgiving, meaning 
that it would resume cooperation if its opponent made just one move in that 
direction.

Afterwards another ecological computer tournament was carried out by Nowak and 
Sigmund \cite{ns93}, where the initial competing strategies
were selected as follows. They described a strategy by four conditional 
probabilities: [$p_R$,$p_S$,$p_T$,$p_P$] that determine, respectively, the
probability that the strategy play C after receiving the payoff $R$, $S$, $T$, or $P$. To simulate genetic drift they 
allowed 'mutations'{\it i.e.} the replacement of a given strategy by another random strategy in each round with a small 
probability $p$. In addition they consider a noisy background, parameterized by $\epsilon$ to better model imperfect 
communication in nature. In this simulation, a different strategy was found to be the most stable in
the evolutionary sense. This strategy was approximately [1,0,0,1]. It had previously been named {\it simpleton} by 
Rapoport and Chammah \cite{rc65} and later {\it Pavlov} by mathematicians D. and V. Kraines \cite{kk88},
because if its action results in a high payoff ($T$ or $R$) it stays, but
otherwise it changes its action. Unlike TFT, it cannot invade the 
strategy {\it All D}, given by [0,0,0,0], and like GTFT (Generous TFT, an
strategy that is close to TFT but has an appreciable value of $p_S$) it
is tolerant of mistakes in communication. The main advantage of this 
'win-stay lose-shift' strategy is that it cannot be invaded by a 
gradual drift of organisms close to {\it All D}, unlike TFT, 
since after a single mistake in communication Pavlov is tempted to defect 
($p_T=0$) and will exploit the generous co-operator. This keeps the
population resistant to attack by All D. 

On the other hand, the spatial structure by itself has also been identified
as sufficient element to build cooperation. That is, unconditional players 
(who always play C or D no matter what their opponents play) without 
memory and no strategical elaboration can attain sustainable cooperation 
when placed in a two dimensional array and playing only against their nearest 
neighbors \cite{nm92}.
 
The combination of the two above elements that are known to promote 
cooperation, iterated competition of different conditional strategies and 
spatial structure, was first studied in \cite{ln94} using $m$-step 
memory strategies. 
Later on, Brauchli {\it et al} \cite{bkd99} studied the strategy space 
of all stochastic strategies with one-step memory. 
Here, our approach is in a similar vein: we have a cellular automata (CA)
and attached to each cell a strategy, specified by a 4-tuple of 
conditional probabilities, that dictates how to play (C or D) against its 
neighbors. However, in order to provide a greater "microscopic" insight 
than just the four average values of the conditional probabilities 
(as is the case when 
continuous real conditional probabilities are considered), we resort to 
{\it binary Markovian strategies} (BMS). That is,
conditional probabilities $p_X$ of playing C after getting the 
payoff $X$ that are either 0 or 1, instead of real. There are thus only 
$2^4$=16 possible BMS whose frequencies can be measured. 
Another simplification we introduced is that at each time step a given 
agent plays the same action (C or D) against all its neighbors and 
takes account only its total payoff, collected by playing against them, 
instead of keeping track of each individual payoff. Then depending
if its neighborhood was "cooperative" or not (different possibilities 
to assess this are proposed in section 3) and what it played, it
adopts action C or D  against all its neighbors. 
Therefore, we have a CA such that each cell has a given state which is 
updated taking into account both this state and the {\it outer total} 
corresponding to the rest of the neighborhood. Or in the language of 
cellular automata, an {\it i.e. outer totalistic} CA \cite{wolf84}.
We choose totalistic CA because, besides their simplicity and 
their known properties of symmetry \cite{NKS}, 
the results show greater robustness being less dependent on the 
initial configuration. 

Besides deterministic automata, we explore two sources of stochastic behavior: 
Firstly in the update rule, specifically in the criterion to assess if the neighborhood is cooperating 
or not. Secondly, by introducing a (small) noise parameter $\epsilon$ and replacing the values of the conditional probabilities $p_X$, 
0 or 1, by $\epsilon$, 1-$\epsilon$, respectivelly.

In summary, we analyze the evolution in the strategy space that occur for 
different:

\begin{itemize} 

\item $2 \times 2$ games {\it i.e.} different regions in the parameters 
space. 

\item Types of neighborhood: von Neumann and Moore neighborhoods.

\item Update rules: deterministic and stochastic.  

\item  Amounts of noise, measured by a parameter $\epsilon$.

\end{itemize}

This paper is organized as follows.
We begin in section 2 by briefly reviewing some useful 2 $\times$ 
2 games in Biology and Social Sciences. We then present a two 
entries 16 $\times$ 16 table for the pairwise confrontation of BMS, whose 
cells represent the asymptotic (after a transient) average payoff of row 
strategy when playing against the column one.
In section 3, we describe our model and its variants.
Next, in section 4, we present the main results. 
Finally, section 5 is devoted to discussion and final remarks.

\section{The strategic tournament between Markovian strategies in 
$2 \times 2$ non-zero sum games}

A change in the rank order of the 4 payoffs gives rise to games different
from the PD. Some of them are well studied games in biological or social
sciences contexts. We will comment on some of them. 
\begin{center}
\begin{figure}[ht]
\includegraphics[width=0.9\textwidth, height=!]{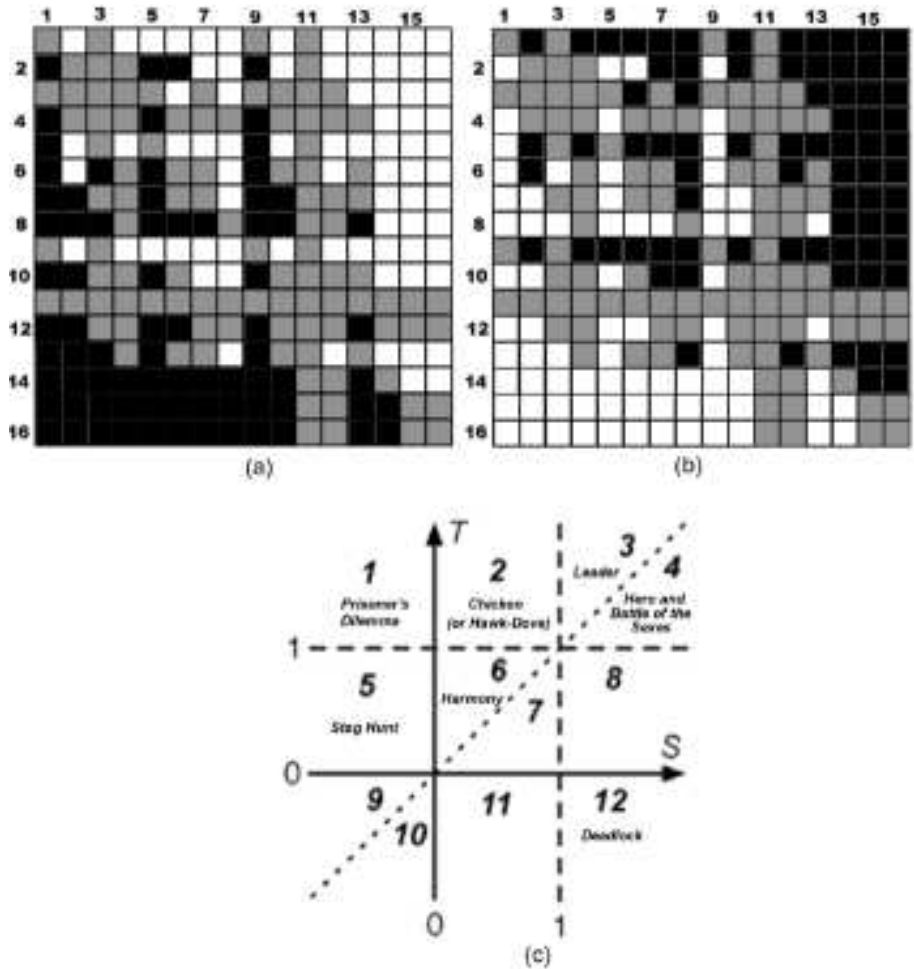}
\caption{Winners matrix - 16 $\times$ 16 strategies.
(a): Games with $T>S$, Prisoner's Dilemma, Chicken, etc; 
(b): Games with $T<S$, Hero and Battle of Sexes, Deadlock game, etc.  
Color coding: White = row wins over column, Black 
(inverse) and Gray = tie. The number reference for each of the 16
possible binary 4-tuples, $[p_R,p_S,p_T,p_P]$ is given by binary number
represented by the 4-tuple plus 1, {\it i.e.}: No. of strategy $=
8p_R+4p_S+2p_T+p_P+1$;
(c)the 12 different possible $2 \times 2$ games marked as zones 
in the parameters space.} 
\label{table} 
\end{figure} 
\end{center}
For instance, when the damage from mutual defection in the PD is increased
so that it finally exceeds the damage suffered by being exploited:
$T>R>S>P$ the new game is called the {\it chicken} or {\it Hawk-Dove} 
(H-D) game. Chicken is named after the car racing game. Two cars drive
towards each other for an apparent head-on collision. Each 
player can swerve to avoid the crash (cooperate) or keep going (defect).
This game applies thus to situations such that mutual defection is the 
worst possible outcome (hence an unstable equilibrium).
The 'Hawk' and 'Dove' allude to the two alternative behaviors displayed by
animals in confrontation: hawks are expected to fight for a resource and will
injure or kill their opponents, doves, on the other hand, only bluff  
and do not engage in fights to the death. So an encounter between two
hawks, in general, produce the worst payoff for both.  

When the reward of mutual cooperation in the chicken game is decreased so
that it finally drops below the losses from being exploited:
$T>S>R>P$ it transforms into the {\it Leader} game. The name of the game 
stems from the following every day life situation: Two car drivers want to 
enter a crowded one-way road from opposite sides, if a small gap occurs in 
the line of the passing cars, it is preferable that one of them take the 
lead and enter into the gap instead of that both wait until a large gap 
occurs and allows both to enter simultaneously.
When $S$ in the Leader game increases so that it finally surpasses 
the temptation to defect {\it i.e.} $S>T>R>P$ the game becomes the {\it Hero} 
game alluding to an "Heroic" partner that plays C against a 
non-cooperative one.  
 
Finally, a nowadays popular game in social sciences is the {\it Stag Hunt}
game, corresponding to the payoffs rank order $R>T>P>S$ {\it i.e.} when the
reward $R$ for mutual cooperation in the PD games surpasses the temptation
$T$ to defect. 
The name of the game derives from a metaphor invented by the
French philosopher Jean Jacques Rousseau:
Two hunters can either jointly hunt a stag or individually hunt a rabbit. 
Hunting stags is quite challenging and requires mutual
cooperation. If either hunts a stag alone, the chance of success is minimal.
Hunting stags is most beneficial for society but requires a lot of trust
among its members. 

Figure \ref{table} (c) reproduces the plot of the parameter space for the 12 
different rank orderings of $2 \times 2$ games with R = 1, P = 0, from ref. 
\cite{hauert}. 
Each game refers to a region in the S, T-plane depicted: 
1 Prisoner's Dilemma; 2 Chicken, Hawk-Dove or Snowdrift game; 3 Leader; 4 
Battle of the Sexes; 5 Stag Hunt; 6 Harmony; 12 Deadlock; all other regions    
are less interesting and have not been named.

Let us consider now the tournament between BMS, in which each particular BMS 
plays repeatedly against all the BMS. We then number the 16 strategies 
from 1 to 16 as follows. We asign to the binary 4-tuple $[p_R,p_S,p_T,p_P]$, 
specifying a strategy, the corresponding binary number $\#$
represented by this 4-tuple plus 1, {\it i.e.}  $\#$ $=
8p_R+4p_S+2p_T+p_P+1$. 
It turns out that the repeated game between any pair of strategies is 
cyclic: after some few rounds both strategies come back to their 
original moves. For example, suppose strategy $\#$ 3 ([0,0,1,0])
playing against strategy $\#$ 9 ([1,0,0,1]). The starting movements are 
irrelevant, and let's choose $\#$ 3 playing C and $\#$ 9 playing D. 
The sequence of movements would then be: 
[C,D] $\rightarrow$ [D,D] $\rightarrow$ [D,C] $\rightarrow$ [C,D] {\it i.e.}
we recover the initial state after 3 rounds.
The cycles, in these 16 $\times$ 16/2 confrontations, are either of period
1, 2, 3 or 4. 
Therefore, to compute the average payoffs per round of any pair of  
strategies we have to sum the payoffs over a number of rounds equal 
to the minimum common multiple of  \{1, 2, 3 \& 4\}, 12, and divide by it.
This allows to construct a $16 \times 16$ matrix with the average 
payoffs for row strategy playing against the column one
for an arbitrary set of payoffs $\{ R,T,S,P \}$.

\vspace{2mm}

The average payoffs per round for strategies $i$ and $j$ playing one against 
the other, can be written as 
$u_{ij}=\alpha_{ij} R + \beta_{ij} S + \gamma_{ij} T + \delta_{ij} P$ and 
$u_{ji}=\alpha_{ji} R + \beta_{ji} S + \gamma_{ji} T + \delta_{ji} P$, 
respectively,
where $\alpha_{ij}$ is the probability that strategy $i$ gets the payoff 
$R$, $\beta_{ij}$ is the probability to get the payoff $S$ and so on.  
Because of the symmetries of the payoff matrix M, 
$\alpha_{ij}=\alpha_{ji}$, $\delta_{ij}=\delta_{ji}$, $\beta_{ij}=\gamma_{ji}$
and  $\gamma_{ij}=\beta_{ji}$, since strategies $i$ and $j$ receive $R$ or 
$P$ the same number of times, and $i$ ($j$) receives $T$ when $j$ ($i$) 
receives $S$. 
Hence, the difference $u_{ij}-u_{ji}$ only depends on whether $T$ is below 
or over $S$. As a consequence, the matrix \ref{table}-(a) 
representing the results of the $16 \times 16$ = 256 encounters: 
is the same for all the other games with $T>S$. The same is true 
for \ref{table}-(b) representing the results for all 
the games with $T<S$. In addition note the symmetry between both: one is the 
'negative' of the other.

\begin{table}[h]
\begin{tabular}{|c|c|c|c|c|c|c|}
\hline
\scriptsize{$[p_R,p_S,p_T,p_P]$} & \scriptsize{Asymptotic Average Payoff 
$U$} & \scriptsize{PD} & \scriptsize{Chicken} & 
\scriptsize{Stag Hunt} 
& \scriptsize{Leader} & \scriptsize{Hero} \\
\hline
\scriptsize{$[0,0,0,0]$} & \scriptsize{$8(T+P)$}  & \scriptsize{14.66} & \scriptsize{10.66} & \scriptsize{12.00} & \scriptsize{8.00} & \scriptsize{10.66}\\
\hline 
\scriptsize{$[0,0,0,1]$} & \scriptsize{$(55/24)R+(55/24)S+(41/6)T+(55/12)P$} & \scriptsize{12.00} & \scriptsize{11.00} & \scriptsize{10.67} & 
\scriptsize{9.67} & \scriptsize{11.00} \\
\hline
\scriptsize{$[0,0,1,0]$} & \scriptsize{$(55/24)R+(55/24)S+(55/12)T+(41/6)P$} &\scriptsize{10.33} & \scriptsize{8.33} & \scriptsize{9.67} & 
\scriptsize{7.67} & \scriptsize{8.33} \\
\hline
\scriptsize{$[0,0,1,1]$} & \scriptsize{$4(R+S+T+P)$} & \scriptsize{11.33} & \scriptsize{11.33} & \scriptsize{11.33} & \scriptsize{11.33} & \scriptsize{11.33} \\
\hline
\scriptsize{$[0,1,0,0]$} & \scriptsize{$2S+7T+7P$} & \scriptsize{12.83} & \scriptsize{10.33} & \scriptsize{10.50} & \scriptsize{9.67} & \scriptsize{11.33} \\
\hline 
\scriptsize{$[0,1,0,1]$} & \scriptsize{$(35/12)R+(61/12)S+(61/12)T+(35/12)P$} & \scriptsize{9.67} & \scriptsize{11.17} & \scriptsize{8.67} & 
\scriptsize{12.67} & \scriptsize{12.67} \\
\hline
\scriptsize{$[0,1,1,0]$} & \scriptsize{$(35/12)R+(61/12)S+(35/12)T+(61/12)P$}  & \scriptsize{7.17} & \scriptsize{7.17} & \scriptsize{7.17} & 
\scriptsize{9.67} & \scriptsize{8.67} \\
\hline
\scriptsize{$[0,1,1,1]$} & \scriptsize{$(55/12)R+(41/6)S+(55/24)T+(55/24)P$} & \scriptsize{7.67} & \scriptsize{9.67} & \scriptsize{8.33} & 
\scriptsize{12.00} & \scriptsize{10.67} \\
\hline
\scriptsize{$[1,0,0,0]$} & \scriptsize{$2R+7T+7P$} & \scriptsize{14.83} & \scriptsize{11.33} & \scriptsize{13.17} & \scriptsize{8.00} & \scriptsize{10.33} \\
\hline
\scriptsize{$[1,0,0,1]$} & \scriptsize{$(61/12)R+(35/12)S+(61/12)T+(35/12)P$} & \scriptsize{12.67} & \scriptsize{12.67} & \scriptsize{12.67} & 
\scriptsize{10.17} & \scriptsize{11.17} \\
\hline
\scriptsize{$[1,0,1,0]$} & \scriptsize{$(61/12)R+(35/12)S+(61/12)T+(35/12)P$} & \scriptsize{10.17} & \scriptsize{8.67} & \scriptsize{11.17} & 
\scriptsize{7.17} & \scriptsize{7.17} \\
\hline
\scriptsize{$[1,0,1,1]$} & \scriptsize{$(41/6)R+(55/12)S+(55/24)T+(55/24)P$} & \scriptsize{9.67} & \scriptsize{10.67} & \scriptsize{11.0} & 
\scriptsize{10.33} & \scriptsize{9.67} \\
\hline
\scriptsize{$[1,1,0,0]$} & \scriptsize{$4(R+S+T+P)$} & \scriptsize{11.33} & \scriptsize{11.33} & \scriptsize{11.33} & \scriptsize{11.33} & 
\scriptsize{11.33} \\
\hline
\scriptsize{$[1,1,0,1]$} & \scriptsize{$7R+7S+2T$} & \scriptsize{9.67} & \scriptsize{13.17} & \scriptsize{11.33} & \scriptsize{14.83} & \scriptsize{13.17} \\
\hline
\scriptsize{$[1,1,1,0]$} & \scriptsize{$7R+7S+2P$} & \scriptsize{8.00} & \scriptsize{10.50} & \scriptsize{10.33} & \scriptsize{12.83} & \scriptsize{10.50} \\
\hline
\scriptsize{$[1,1,1,1]$} & \scriptsize{$8(R+S)$} & \scriptsize{8.00} & \scriptsize{12.00} & \scriptsize{10.66} & \scriptsize{14.66} & \scriptsize{12.00} \\
\hline
\end{tabular}
\caption{Asymptotic Average Payoff for Different $2 \times 2$ Games.}
\label{scores} 
\end{table}

For each strategy we calculate $U_i=\sum_j u_{i,j}$ the total (sum over all
the 16 possible contenders) average payoff of strategy $i$. 
The general results of this calculation, as well as the particular numerical
values when the four payoff are \{1.333, 1, 0.5 \& 0\} 
are listed in the table \ref{scores}.
For instance we have the PD game when $R=1$, $T=1.333$, $S=0$ and $P=0.5$;
the Chicken game when  $R=1$, $T=1.333$, $S=0.5$ and $P=0$ and so on.
We observe that for these values of the parameters, $[1,0,0,0]$ is the 
strategy with the highest average payoff for PD and 
Stag Hunt, while $[1,1,0,1]$ is the strategy that has the highest value of 
$V$ for Chicken, Leader and Hero.  

\section{Binary Markovian Strategy Competition in an Outer 
Totalistic Cellular Automata}

Each agent is represented, at time step $t$, by a cell with center at 
$(x,y)$ with a binary {\it behavioral variable} $c(x,y;t)$ that takes 
value 1 (0) if it is in the C (D) state.    
At every time step a given cell plays pairwise a $2 \times 2$ game against all 
of its neighbors collecting total utilities $U(x,y;t)$ given by the sum of 
the payoffs $u(x,y;t)$ it gets against each neighbor.

We use a rescaled payoff matrix in which the 2nd best payoff $X^{2nd}$ is
fixed to 1 and the worst payoff, $X^{4th}$ is fixed to 0.
For example, the PD payoff matrix is described by two 
parameters: $\tilde{T}$, $\tilde{P}$; the {\it chicken} PA
by $\tilde{T}$ and $\tilde{S}$, etc. 

We consider two different neighborhoods $N(x,y)$: 
a) the {\it von Neumann neighborhood} ($q=4$ neighbor
cells, the cell above and below, right and left from a given cell) and
b) the {\it Moore neighborhood} ($q=8$ neighbor cells: von Neumann
neighborhood + diagonals).

In the case of ordinary (non totalistic) CA the way the cell at $(x,y)$ 
plays against its neighbor at $(x',y')$ is determined by a 4-tuple 
[$p_R(x,y;t)$, $p_S(x,y;t)$, $p_T(x,y;t)$, $p_P(x,y;t)$] that are the conditional 
probabilities that it plays C at time $t$ if it got at time 
$t$-1 $u(x,y;t-1)=R, T, S$ or $P$ respectively. 
Here, as we anticipated, we use a totalistic automata and then at each time 
step every cell plays at once a definite action (C or D) against all its $q$ 
(4 or 8) neighbors instead of playing individually against each neighbor. 
Hence, it is necessary to extend the above conditional probabilities in 
such a way that they take into account the neighborhood "collective" state.
In order to do so, note that the conditional probabilities $p_R$, $p_S$
, $p_T$ \& $p_P$ can also be regarded as, respectively, the probability of 
playing C after [C,C], [C,D], [D,C] \& [D,D].   
Then a natural way to extend these conditional probabilities is to consider 
that "the neighborhood plays C (D)" if the majority of its 
neighbors play C 
(D), that is, if 

$$q_C(x,y;t) \equiv \frac{\sum_{N(x,y)} c(x',y',t)}{q},$$

is above or below 1/2.
There are different ways to implement this. Let us 
consider the following two variants, one in terms of a deterministic update 
rule for the behavioral variable, and the other in terms of an stochastic
update rule.  

\begin{itemize}

\item {\it Deterministic} update: 
\begin{equation}
\begin{array}{ll}
c(x,y;t+1)=c(x,y;t)[p_R \theta^+(q_C(x,y;t)-q/2)+p_S \theta^+(q/2-q_C(x,y;t)]+\\
\;\;\;\;\;\;\;\;\;\;\;\;\;\;\;(1-c(x,y;t))[p_T \theta^+(q_C(x,y;t)-q/2)+
p_P \theta^+(q/2-q_C(x,y;t))],
\label{eq:deter}
\end{array}
\end{equation}
where $\theta^+ (q_C(x,y;t))$ is a Haviside step function given by:
\begin{equation}
\theta^+(x)=,\left\{ 
\begin{array}{ll}
 1 & \textrm{if $x>0$}\\
 0  & \textrm{if $x\le 0$}
\end{array}\right.
\end{equation}

\item{\it Stochastic} update:   
\begin{equation}
\begin{array}{ll}
c(x,y;t)=
c(x,y;t-1)q_C(x,y;t-1)p_R(x,y;t)+\\
\;\;\;\;\;\;\;\;\;\;\;\;\;\;\;\;\;\;c(x,y;t)(1-q_C(x,y;t-1)) p_S(x,y;t) +\\
\;\;\;\;\;\;\;\;\;\;\;\;\;\;\;\;\;\; (1-c(x,y;t))q_C(x,y;t)p_T(x,y;t)+ \\
\;\;\;\;\;\;\;\;\;\;\;\;\;\;\;\;\;\;(1-c(x,y;t-1))(1-q_C(x,y;t-1))p_P(x,y;t).
\label{eq:stochup}
\end{array}
\end{equation}
where the probability that the neighborhood plays C is equal to the fraction $q_C(x,y;t)$ of C neighbors.
\end{itemize}

Finally, after updating its behavioral variable, each agent updates its four 
conditional probabilities 
$p_X$ copying the ones of the individual belonging to $\tilde{N}(x,y)$ who
got the maximum utilities.

In order to take into account errors in the behavior of agents we include a
noise level by a parameter $\epsilon>0$ in such a way that the conditional
probabilities $p_X$ for each agent can take either the value 
$\epsilon$ or the value $1-\epsilon$. 

We consider a square network with $N_{ag}=L \times L$ agents and periodic
boundary conditions.
We start from an initial configuration in which each
agent is assigned randomly a strategy in such a way that the 16
possible strategies
are randomly represented among the population of the $N_{ag}$ agents.

The results of this simulations, like the ones in references \cite{bkd99} are 
sensible to random seed selection, so, to avoid such dependence, 
frequencies for the 16 BMS as a function of time are obtained by averaging 
over $N_c$ simulations with different random seeds.  

In the next section we present  the results for several different games.

\section{Results}

The results at this section correspond to 
averages over an ensemble of $N_c=100$ different random initial conditions 
for 100$\times$ 100 lattices\footnote{Results do not change substantially
when the lattice size or the number 
of averaged initial conditions is increased.}. 
For all the games, we study the time evolution of the average frequency 
for each of the 16 strategies {\it i.e.},
the fraction of the $L\times L$ agents in the network that plays with that 
given strategy.

\subsection{Deterministic Prisoner's Dilemma}

\begin{center}
\begin{figure}[ht]
\begin{tabular}{cc}
\includegraphics[width=0.5\textwidth, height=!]{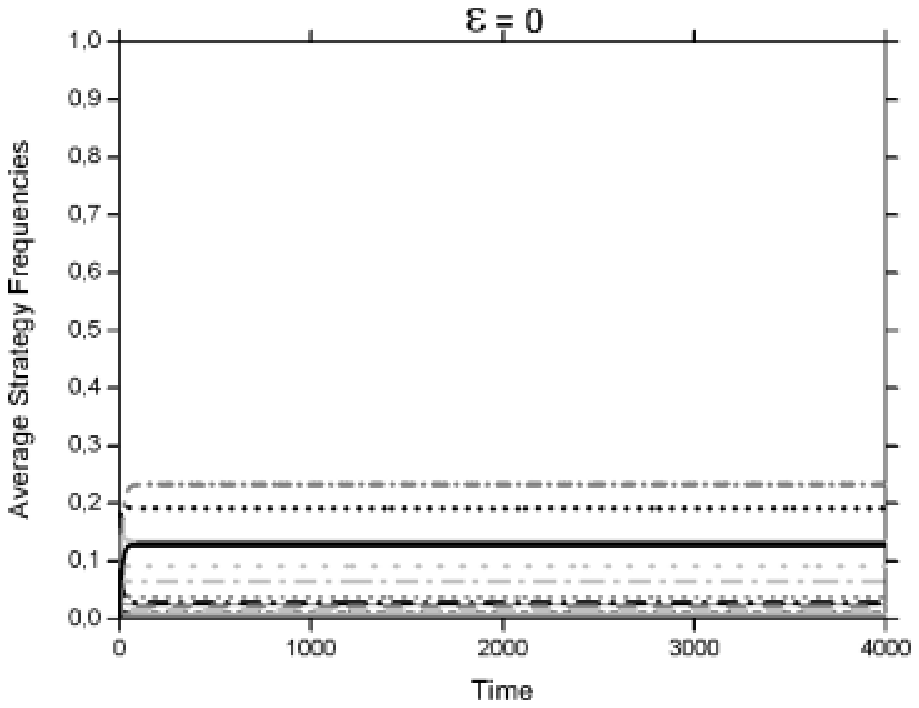} &
\includegraphics[width=0.5\textwidth,height=!]{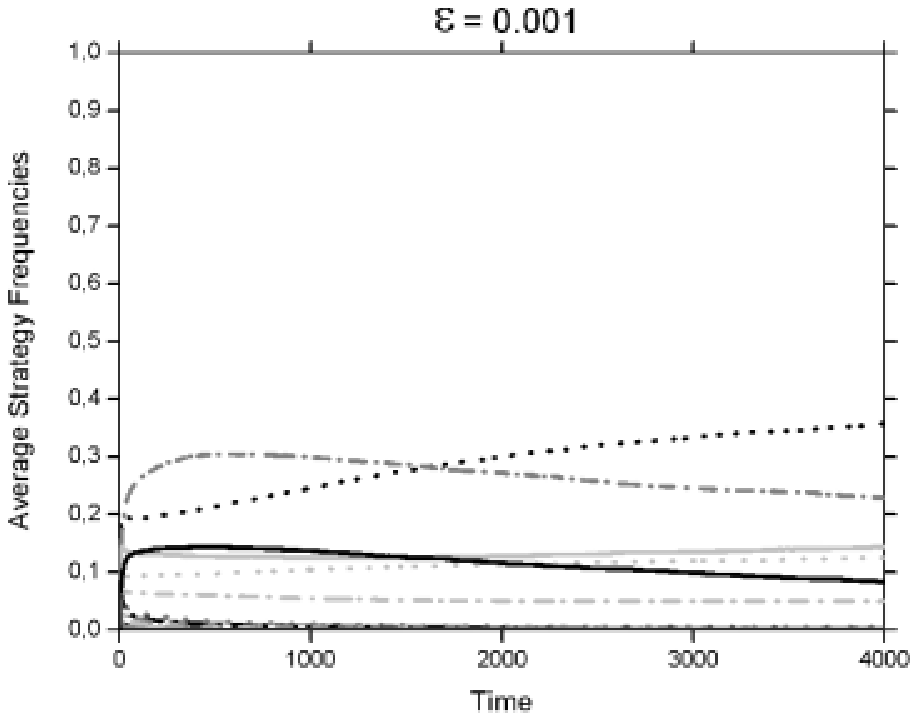} \\
\includegraphics[width=0.5\textwidth,height=!]{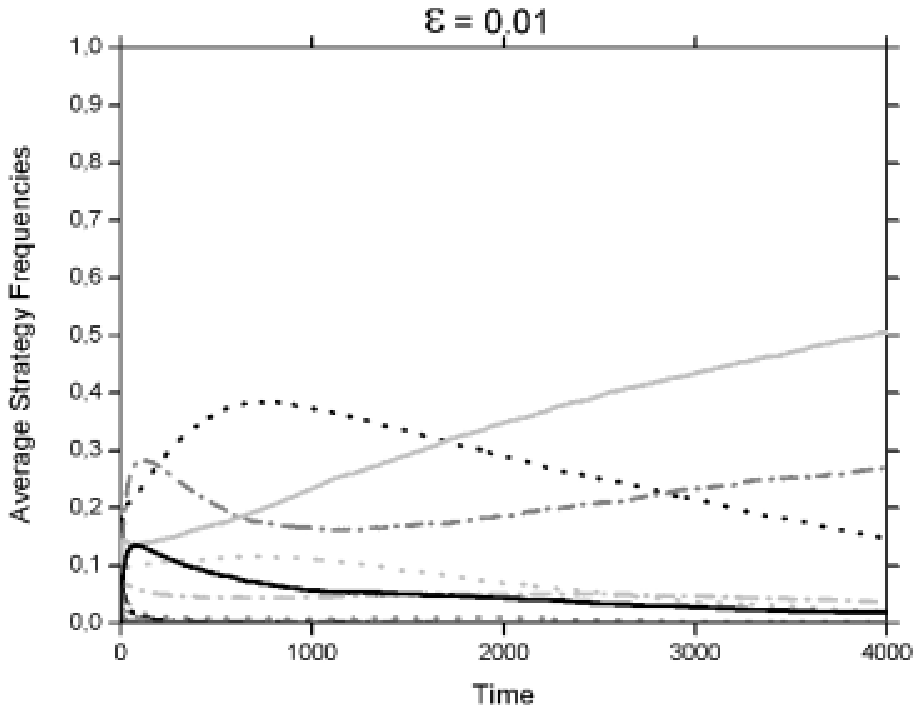} &
\includegraphics[width=0.5\textwidth,height=!]{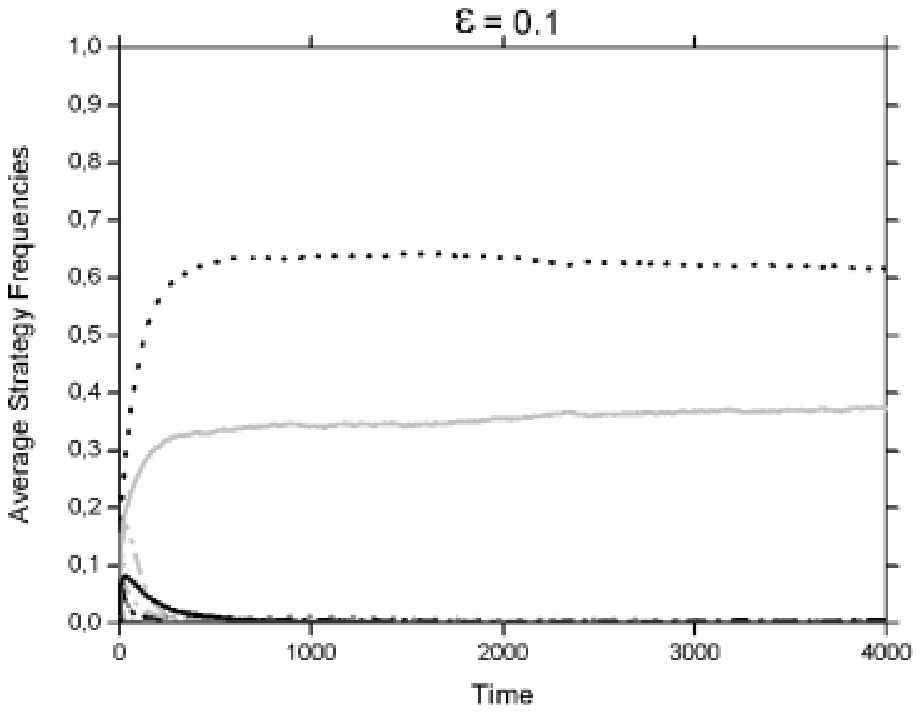}
\end{tabular}
\caption{Frequencies for the 16 competing Binary Markovian Strategies
(BMS) vs the number of time steps for Deterministic Prisoner's Dilemma 
with $\tilde{T}=1.333$, $\tilde{P}=0.5$ and $q=4$. 
The strategy $[p_R,p_S,p_T,p_P]$ references used in all the figures are 
the following: $[0,0,0,0]$=thick light gray dotted curve (ALLWAYS D), 
$[0,0,0,1]$= thin black dashed curve,$ [0,0,1,0]$= thick black dotted 
curve, $[0,0,1,1]$= thick gray dash-dotted curve, $[0,1,0,0]$= thick gray 
dotted curve, $[0,1,0,1]$= thin gray solid curve, $[0,1,1,0]$= thick 
black dash-dotted curve, $[0,1,1,1]$= thin light 
gray solid curve, $[1,0,0,0]$= thick light gray dash-dotted curve, 
$[1,0,0,1]$=thick black dashed curve (PAVLOV), $[1,0,1,0]$=thick light 
gray solid curve (TFT), $[1,0,1,1]$=thick black solid curve, $[1,1,0,0]$= 
thick gray dashed curve, $[1,1,0,1]$=thick light gray dashed curve, 
$[1,1,1,0]$= thick gray solid curve, and $[1,1,1,1]$= thin black solid 
curve (ALLWAYS C).   }
\label{fig:z4det4ep}
\end{figure}
\end{center}

First we see the deterministic PD for $\tilde{T}=1.333$
and $\tilde{P}=0.5$. 
Fig.\ref{fig:z4det4ep} shows the frequencies for
the 16 different BMS, for the $q=4$ von Neumann neighborhood.
One can see that without noise ($\epsilon=0$) the system reaches
quickly a dynamic equilibrium state in which
several of the 16 strategies are present.   
As long as $\epsilon$ grows, the number of strategies decreases, and the
diversity 
without noise transforms into two surviving strategies: [0,0,1,0] and TFT
([1,0,1,0]). 
Indeed, the results 
for $\epsilon=0.1$ correspond to some cases in which 
the whole population ends with the strategy [0,0,1,0] and cases where the 
whole population ends using TFT. But without coexistence of both strategies.
This also explains the lack of fluctuations in the 
results despite of the large value of the noise parameter.

\begin{center}
\begin{figure}[ht]
\begin{tabular}{cc}
\includegraphics[width=0.5\textwidth, height=!]{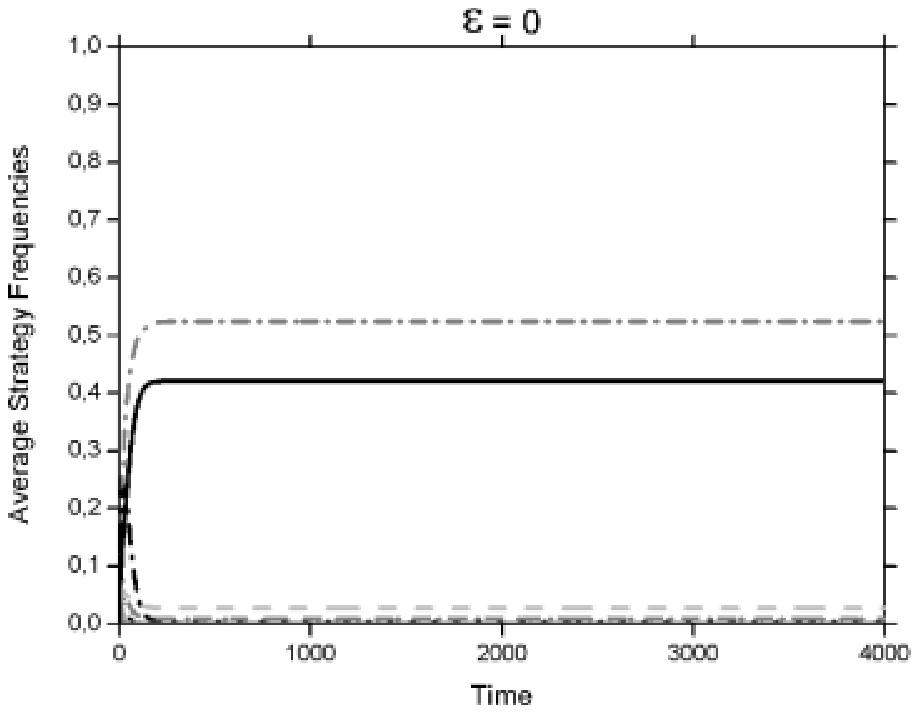} &
\includegraphics[width=0.5\textwidth,height=!]{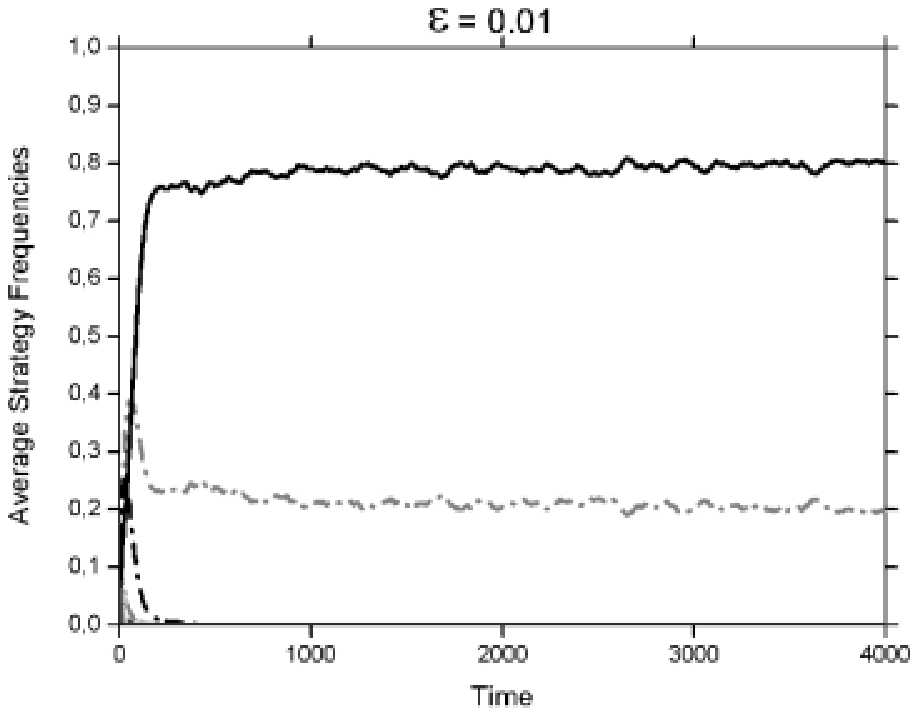} \\
\includegraphics[width=0.5\textwidth,height=!]{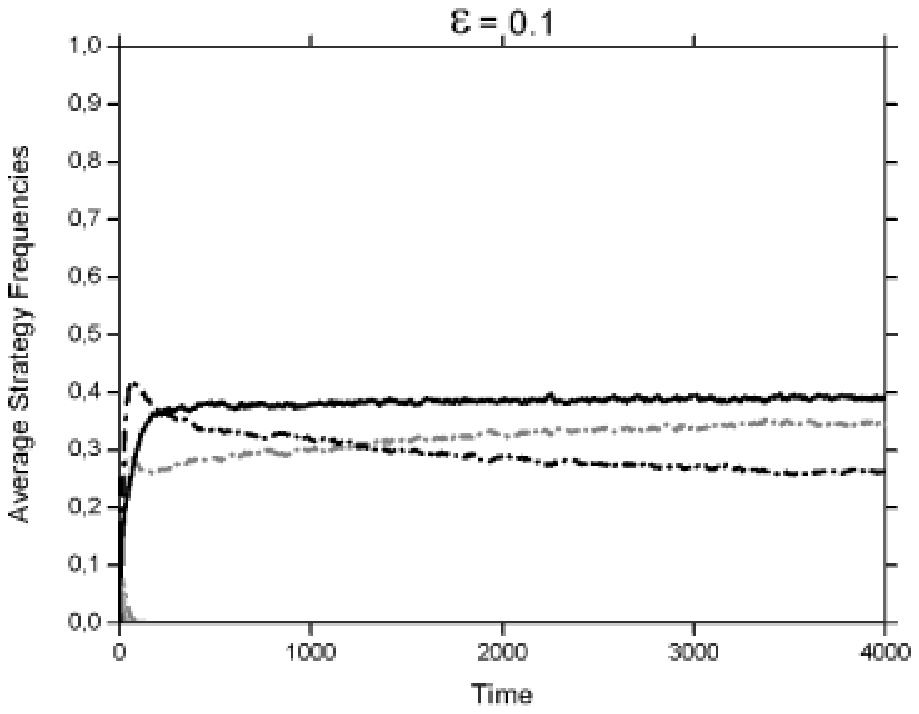} &
\includegraphics[width=0.5\textwidth,height=!]{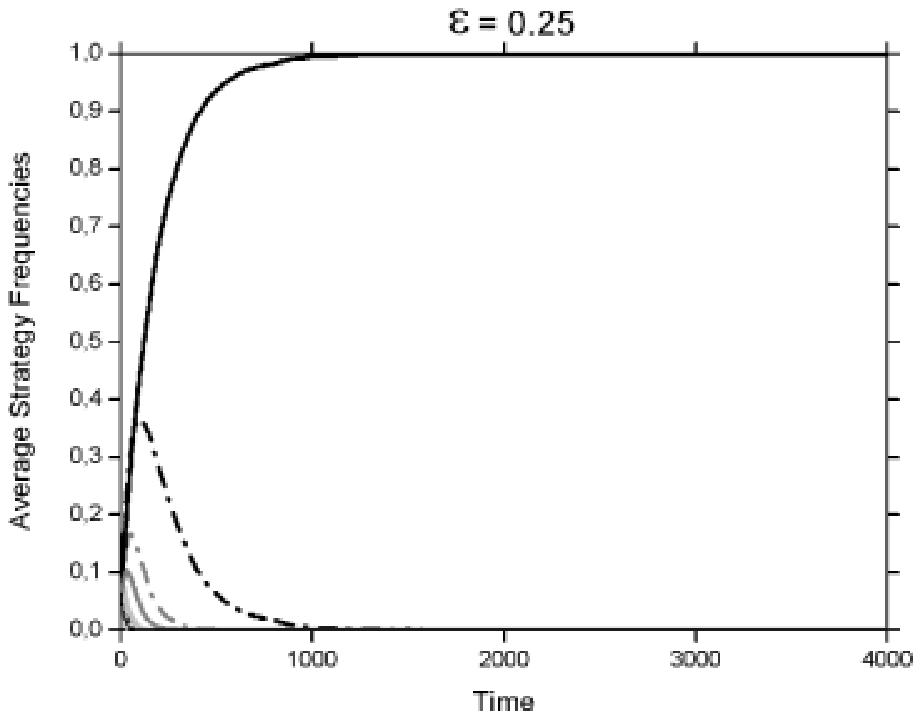}
\end{tabular}
\caption{Frequencies for the 16 competing BMS vs the number of time steps
for Deterministic Prisoner's Dilemma with $\tilde{T}=1.333$, $\tilde{P}=0.5$ 
and $q=8$.
Color and line style codes are the same than in Figure \ref{fig:z4det4ep}}
\label{fig:z8det4ep}
\end{figure}
\end{center}

Results for the Moore neighborhood, $q=8$, are different from
those obtained for the $q=4$, as can be seen from
Fig. \ref{fig:z8det4ep}.

Notice that for zero or a small noise amount of noise 
($\epsilon \le 0.01$)  we have two surviving strategies [0,0,1,1] and
[1,0,1,1]. For $\epsilon=0.1$ a new competing strategy, [0,1,1,0] appears, 
and agents 
distribute almost equally among this strategy, [1,0,1,1], and [0,0,1,1].
Finally, for $\epsilon=0.25$, 100% of the agents follow [1,0,1,1].
\subsection{Stochastic Version}

\begin{center}
\begin{figure}[ht]
\begin{tabular}{cc}
\includegraphics[width=0.5\textwidth, height=!]{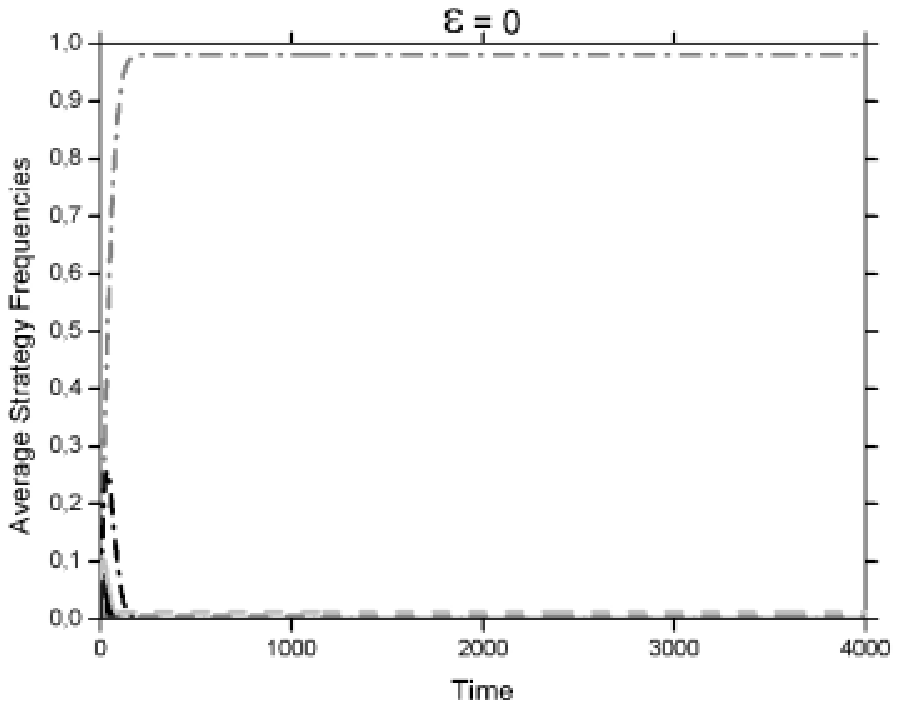} &
\includegraphics[width=0.5\textwidth,height=!]{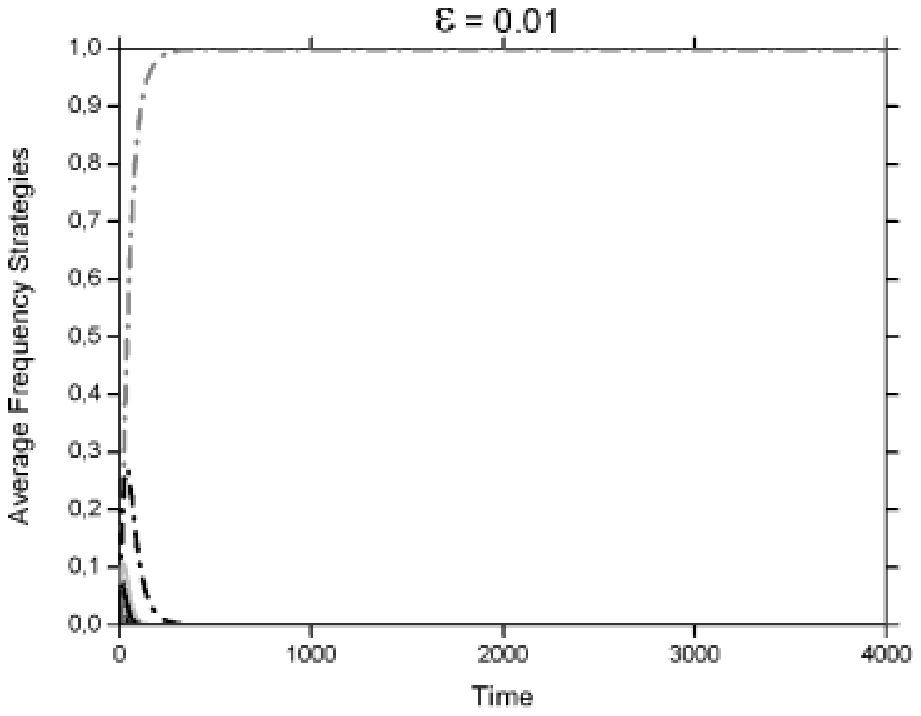} \\
\includegraphics[width=0.5\textwidth,height=!]{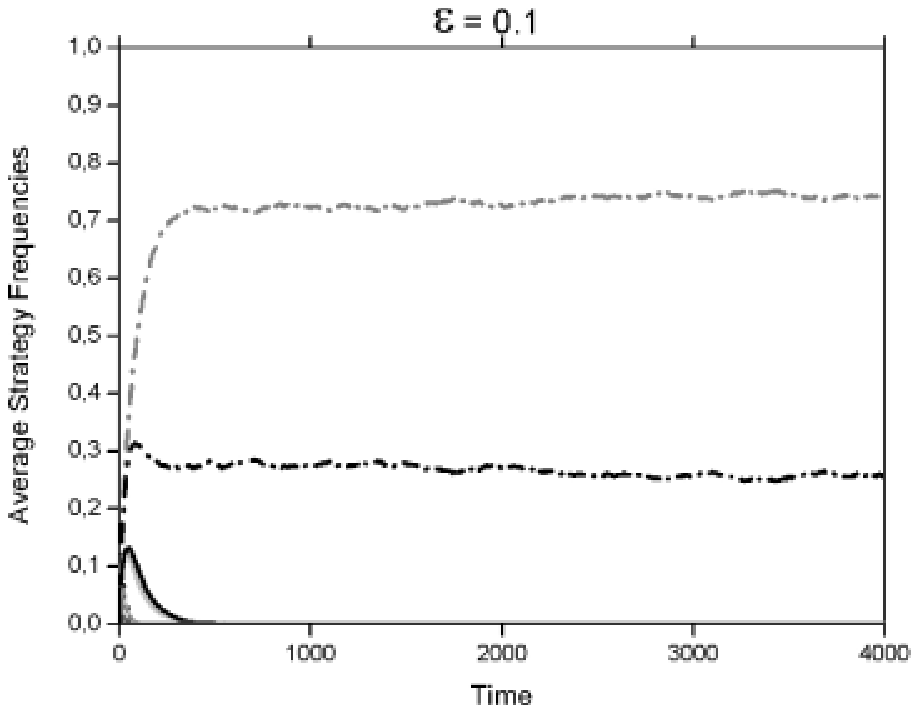} &
\includegraphics[width=0.5\textwidth,height=!]{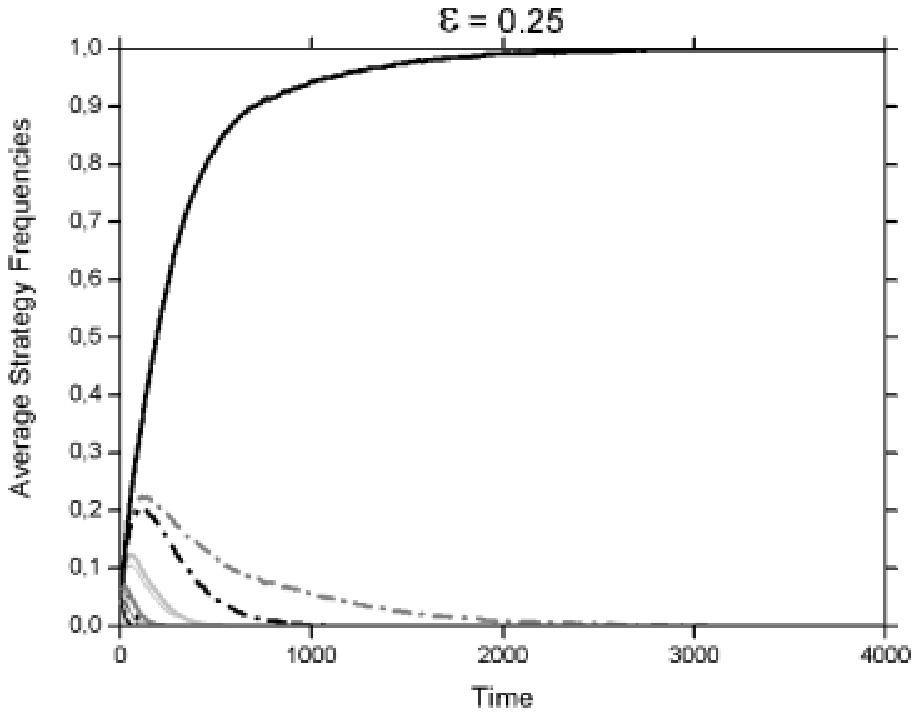}
\end{tabular}
\caption{Frequencies for the 16 competing BMS vs the number of time steps
for Stochastic Prisoner's Dilemma with $\tilde{T}=1.333$, $\tilde{P}=0.5$ 
and $q=8$.
Color and line style codes are the same than in Figure \ref{fig:z4det4ep}.}
\label{fig:z8stoch4ep}
\end{figure}
\end{center}

For the stochastic version higher levels of noise (larger values of 
$\epsilon$) are needed in
order to measure departures from the 0 noise situation. This is natural 
since there is an intrinsic stochastic component in this case. 
Fig.\ref{fig:z8stoch4ep} is a plot of each of the 16 frequencies vs time
for the PD Stochastic game with $q=8$. For zero or small noise 
{\it i.e.} $\epsilon \le 0.01$ the only strategy present is [0,0,1,1]. For
$\epsilon=0.1$ [0,0,1,1] is still the more abundant strategy, but
now it coexists with [0,1,1,0]. If the noise level is increased even more,
the frequency of strategy [1,0,1,1] starts to become non negligible, till
for $\epsilon=0.25$ it controls 100\% of the population.

The average winning strategies are robust for both the stochastic and the 
deterministic case 
with respect to the parameters $\tilde{T}, \tilde{P}$ of PD payoff matrix,
even for 
$\tilde{T}=2$, as long as we are in the region of the Prisoner's Dilemma 
game the behavior is qualitatively the same.

\subsection{Other Payoff Matrices}

In this section we explore other games for the deterministic case. 
Let's observe first the effect of permuting the punishment and the sucker's
payoff {\it i.e.} tacking $\tilde{S}$ = 0.5 $>  \tilde{P}$ = 0 
(Chicken game).
\begin{center}
\begin{figure}[ht]
\begin{tabular}{cc}
\includegraphics[width=0.5\textwidth, height=!]{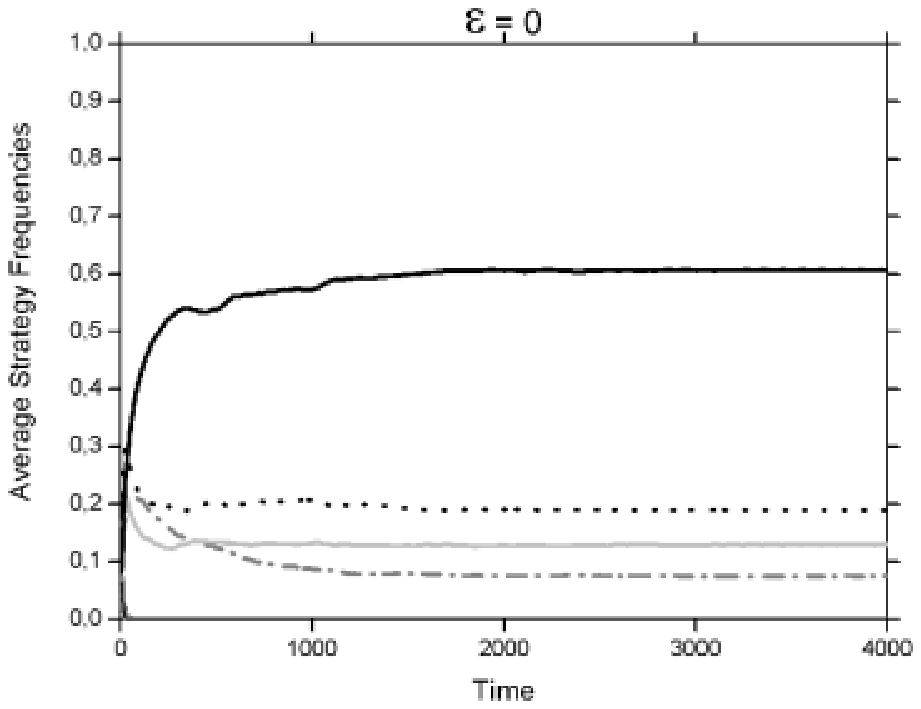} &
\includegraphics[width=0.5\textwidth,height=!]{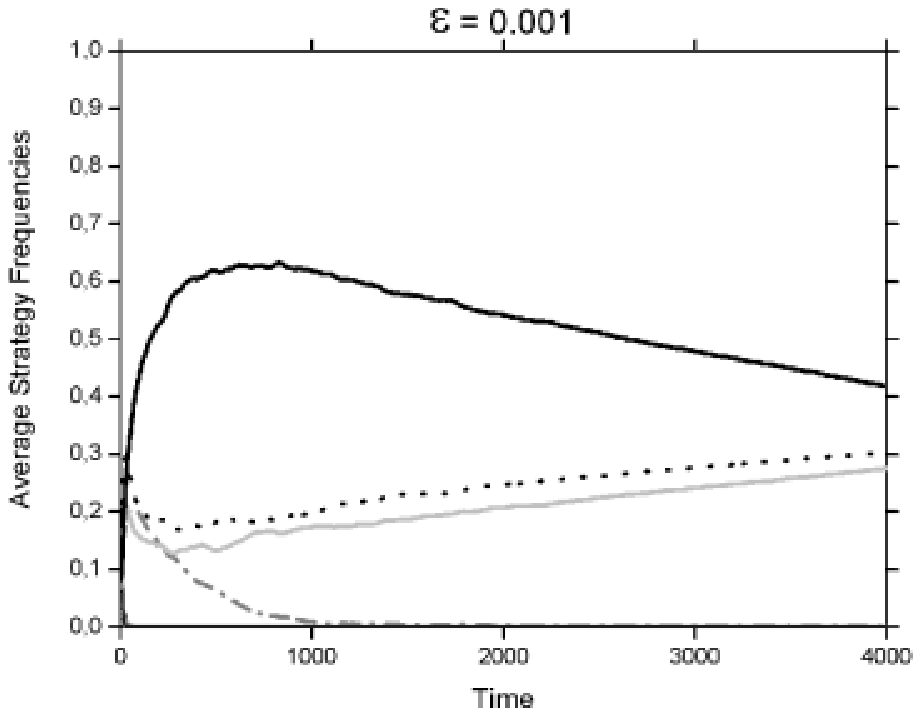} \\
\includegraphics[width=0.5\textwidth, height=!]{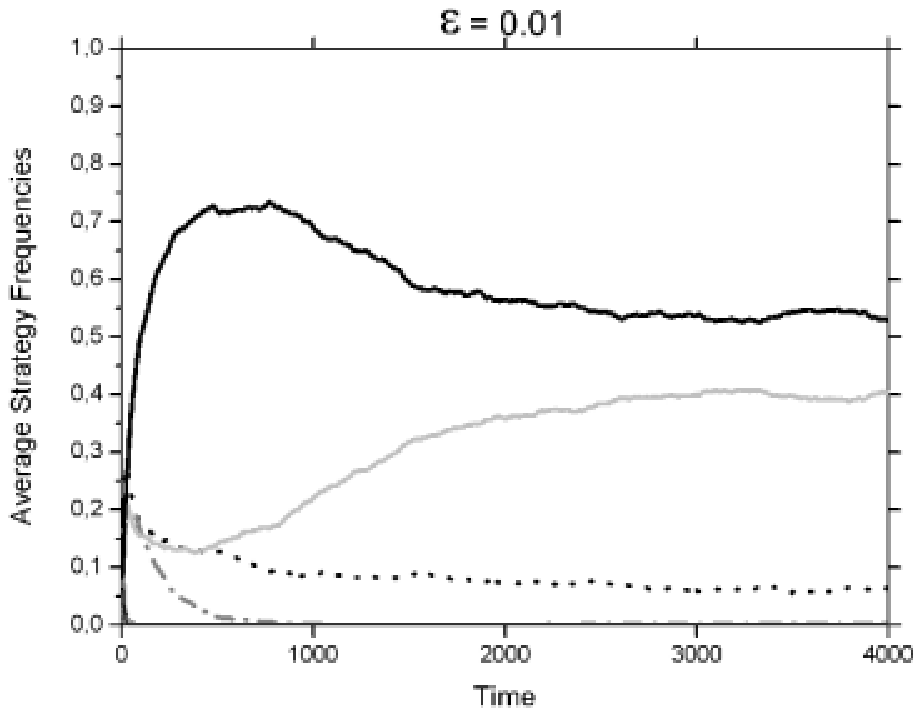} &
\includegraphics[width=0.5\textwidth,height=!]{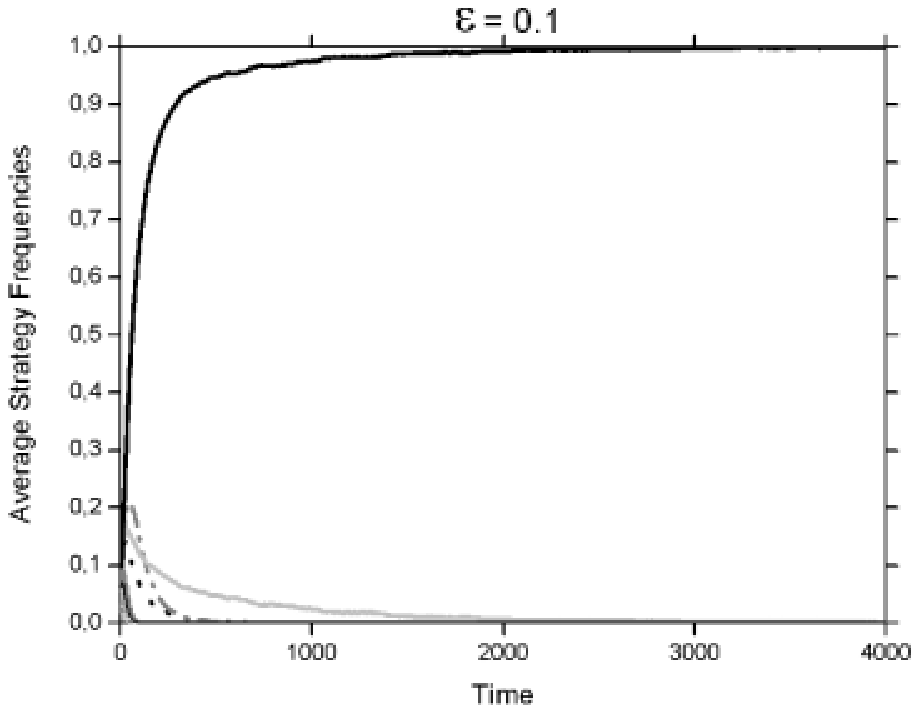}
\end{tabular}
\caption{Frequencies for the 16 competing BMS vs the number of time steps
for Deterministic Chicken Game with $\tilde{T}=1.333$, $\tilde{S}=0.5$ and 
$q=4$.
Color and line style codes are the same than in Figure \ref{fig:z4det4ep}.}
\label{fig:z4chick}
\end{figure}
\end{center}
Results for $q=4$ are plotted in Fig. (\ref{fig:z4chick}). 
For all the considered values of values of $\epsilon$ between 0 and 0.1, 
[1,0,1,1] is the dominant strategy.
For small amounts of noise, coexistence of 3 strategies: 
[1,0,1,1], [0,0,1,0] and TFT ( [1,0,1,0] ). 
Finally, for $\epsilon=0.1$, strategy [1,0,1,1] turns 
to be completely dominant with a frequency of 100 \%.

\begin{center}
\begin{figure}[ht]
\begin{tabular}{cc}
\includegraphics[width=0.5\textwidth, height=!]{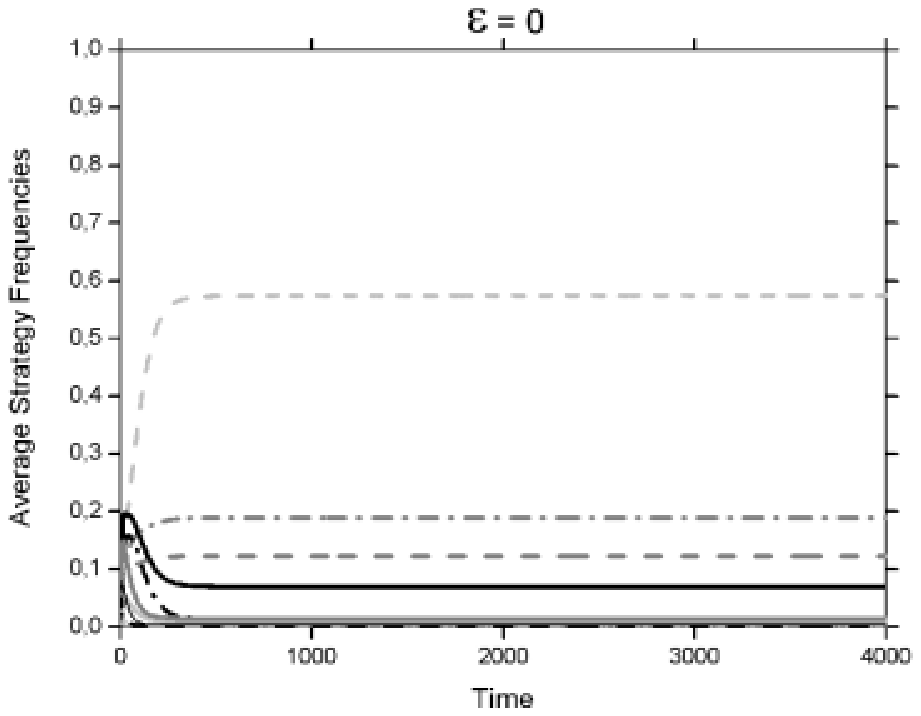} &
\includegraphics[width=0.5\textwidth,height=!]{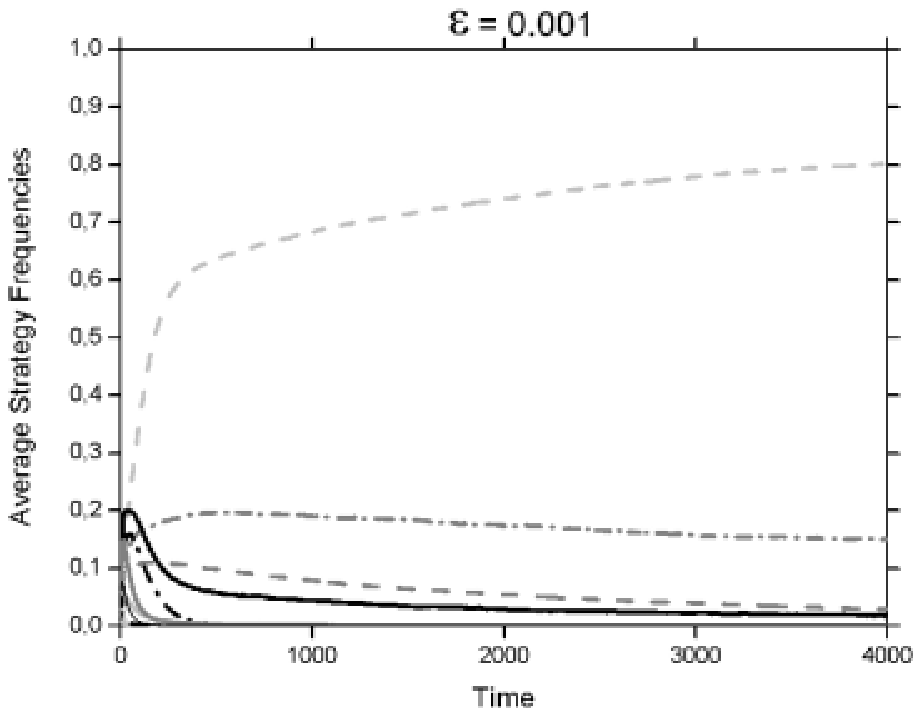} \\
\includegraphics[width=0.5\textwidth, height=!]{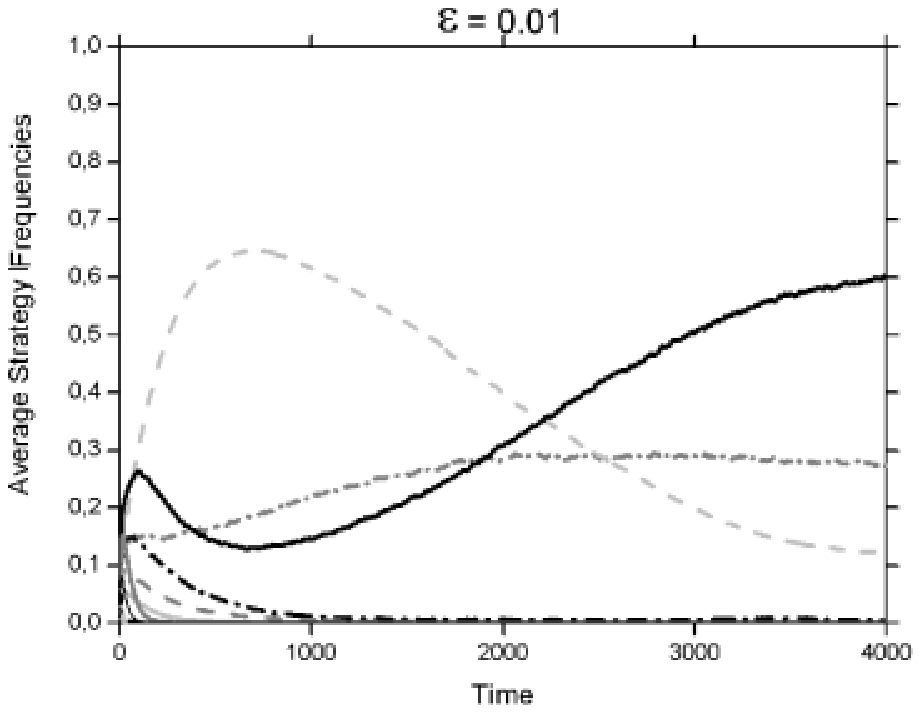} &
\includegraphics[width=0.5\textwidth,height=!]{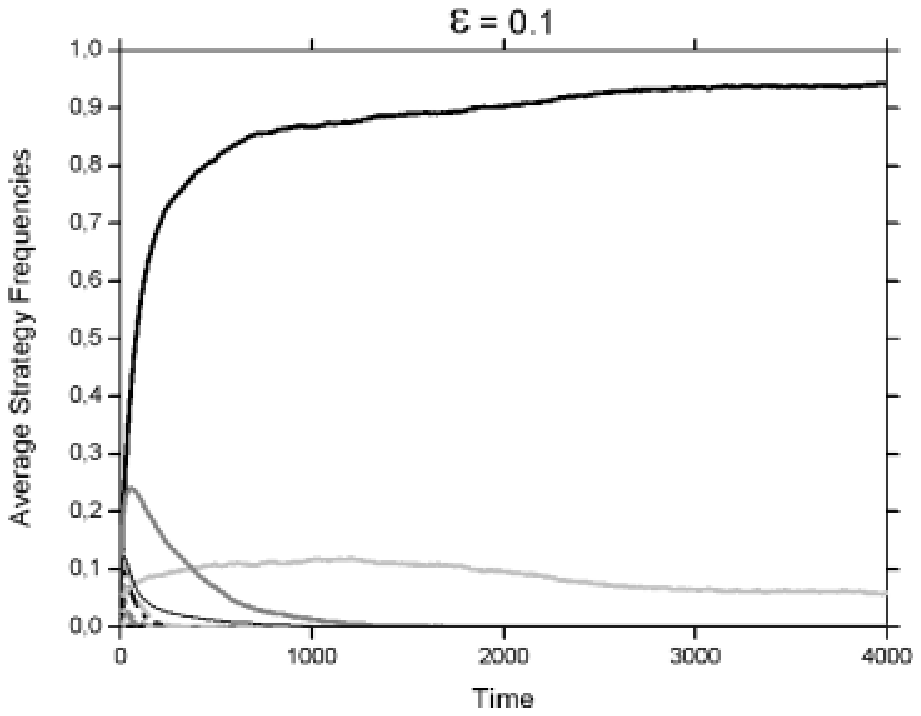}
\end{tabular}
\caption{Frequencies for the 16 competing BMS vs the number of time steps
for Deterministic Chicken Game with $\tilde{T}=1.333$, $\tilde{S}=0.5$, 
and $q=8$. 
Color and line style codes are the same than in Figure \ref{fig:z4det4ep}.
}
\label{fig:z8chick}
\end{figure}
\end{center}
The steady state results for $q=8$ are different for small amounts of noise 
but become qualitatively the same for moderates values of the noise parameter 
($\epsilon \simeq 0.01$), as can be seen from 
Fig. \ref{fig:z8chick}.  

\begin{center}
\begin{figure}[ht]
\begin{tabular}{cc}
\includegraphics[width=0.5\textwidth, height=!]{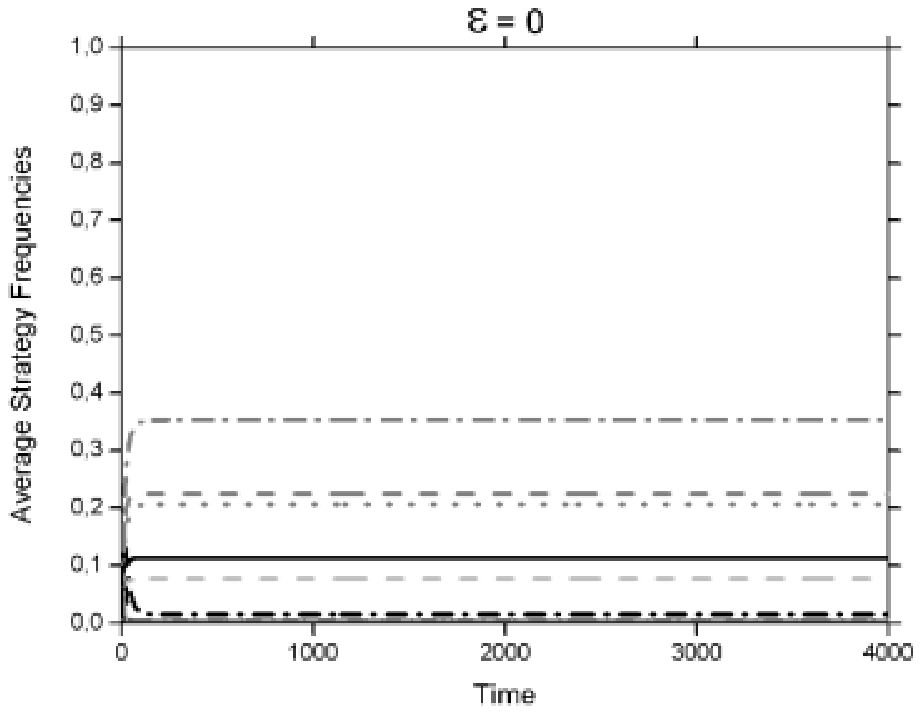} &
\includegraphics[width=0.5\textwidth,height=!]{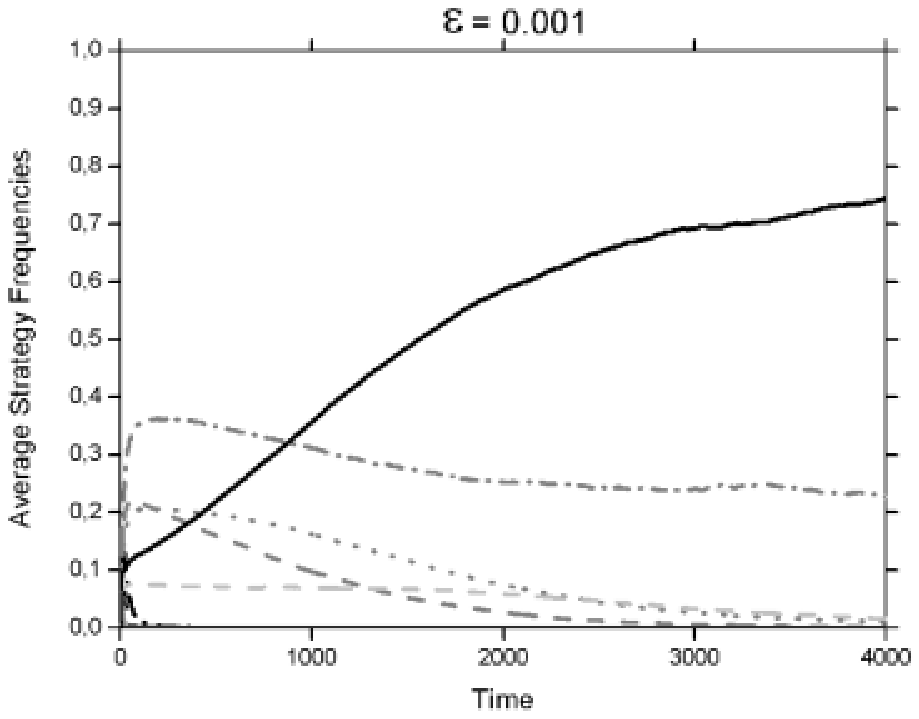} \\
\includegraphics[width=0.5\textwidth, height=!]{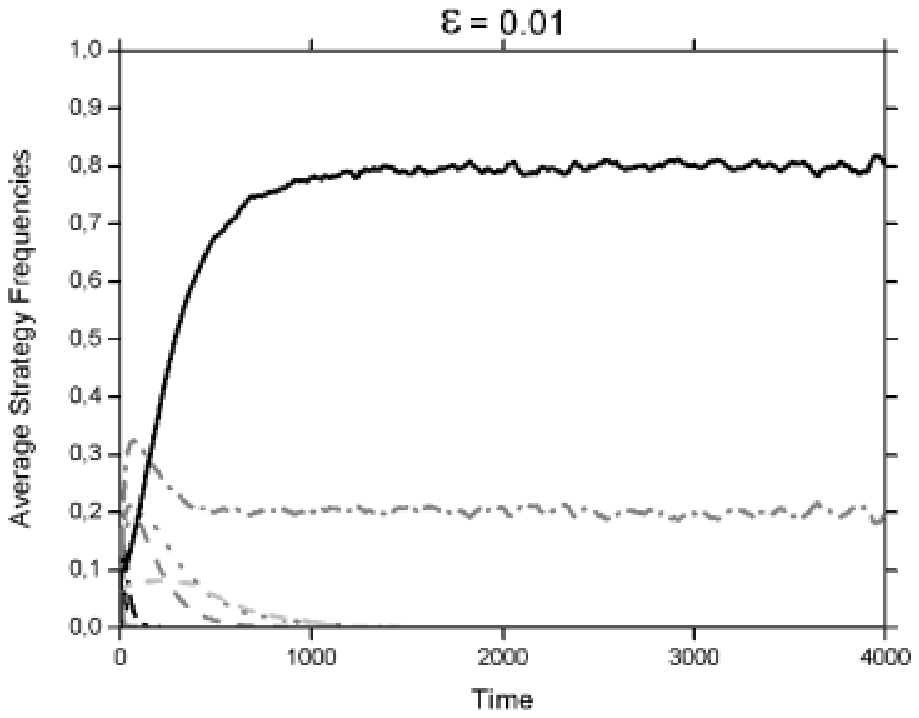} &
\includegraphics[width=0.5\textwidth,height=!]{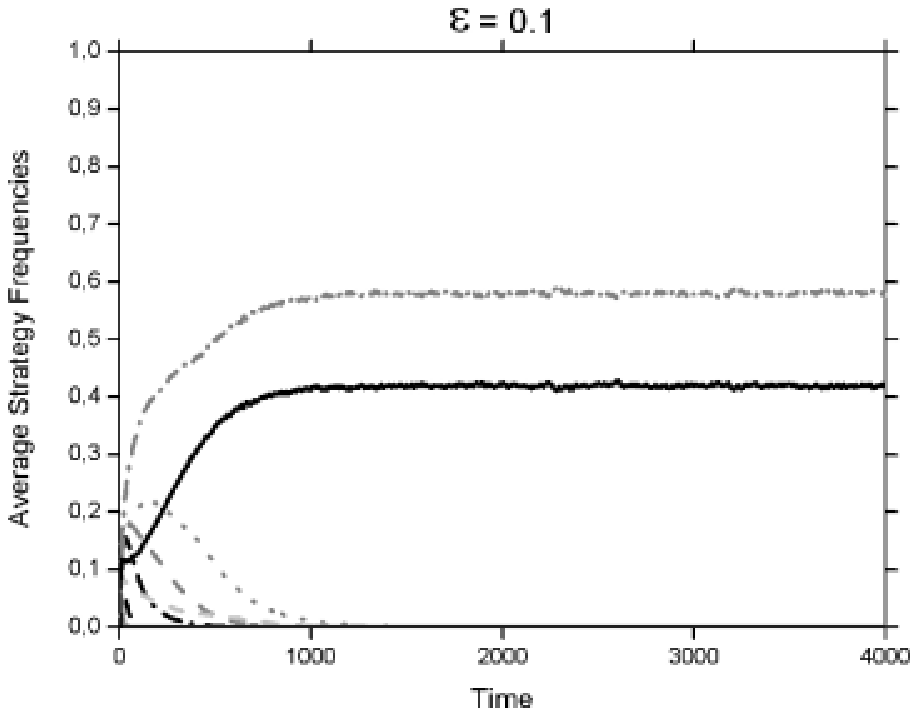}
\end{tabular}
\caption{Frequencies for the 16 competing BMS vs the number of time steps
for Deterministic Leader Game with $\tilde{T}=1.333$, $\tilde{R}=0.5$ and $q=8$.
Color and line style codes are the same than in Figure \ref{fig:z4det4ep}.
}
\label{fig:z8leader}
\end{figure}
\end{center}

The results for the Leader deterministic game 
with $q=8$ are plotted in Fig. \ref {fig:z8leader}.
For the case without noise ($\epsilon=0$) we have a remarkable diversity of 
surviving strategies. 
As $\epsilon$ grows this diversity transforms into only two surviving 
strategies: [1,0,1,1] and [0,0,1,1], whose relative 
dominace is exchanged for the large values of noise ($\epsilon=0.1$).  
Notice that the lack of random fluctuations when $\epsilon=0.1$ 
(a relatively high noise parameter) in figure 
\ref {fig:z8leader} is explained as before because 
the averages correspond either to cases in which 
the whole population selected [0,0,1,1] 
or [1,0,1,1] as their strategies, {\it i.e.}:  there are no coexisting  
strategies in the steady state.

\begin{center}
\begin{figure}[ht]
\begin{tabular}{cc}
\includegraphics[width=0.5\textwidth, height=!]{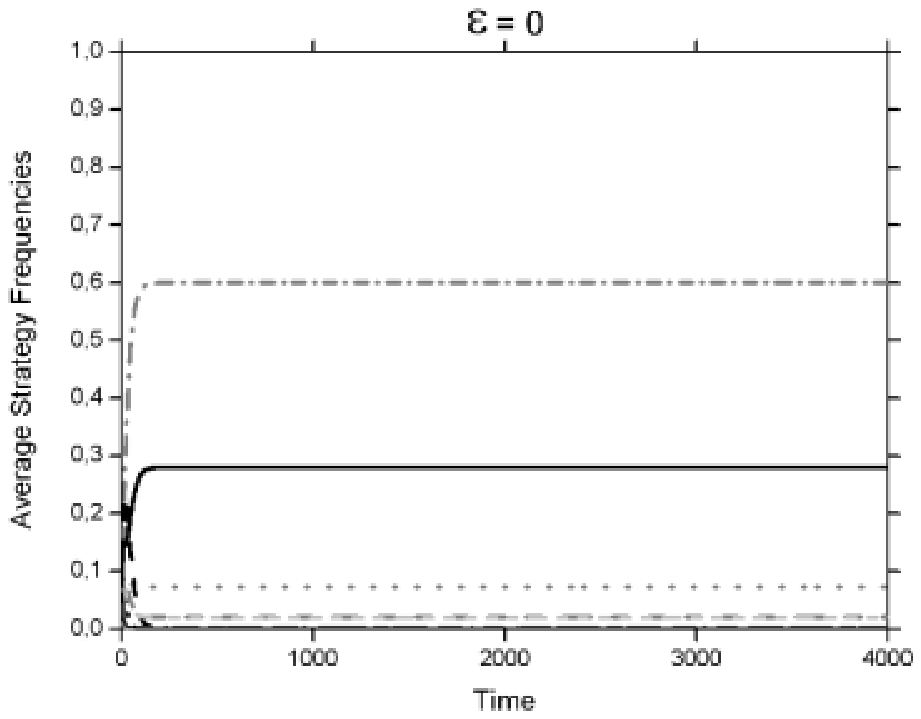} &
\includegraphics[width=0.5\textwidth,height=!]{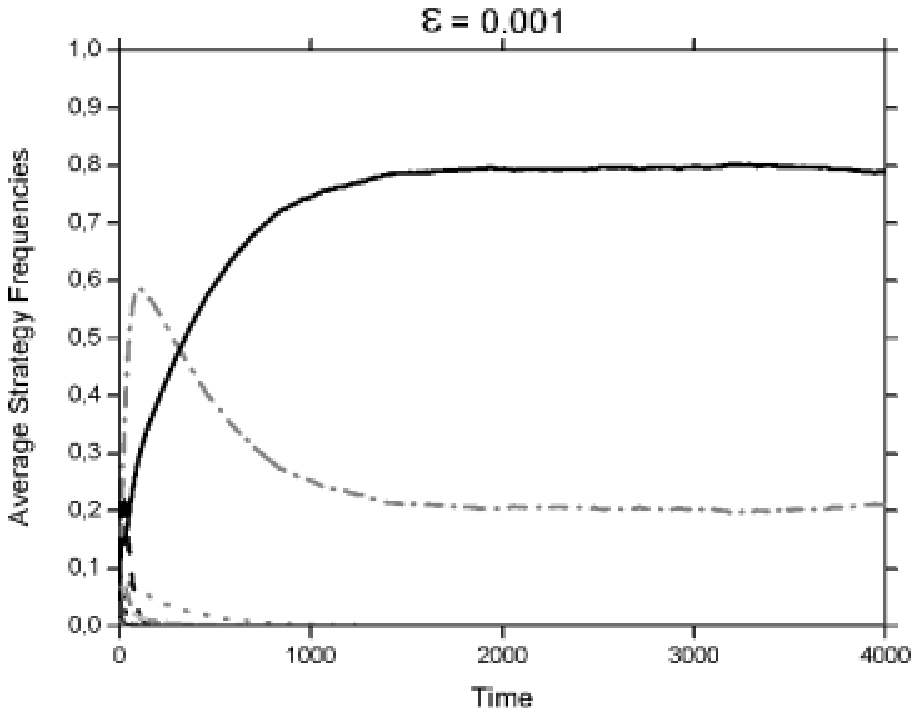} \\
\includegraphics[width=0.5\textwidth, height=!]{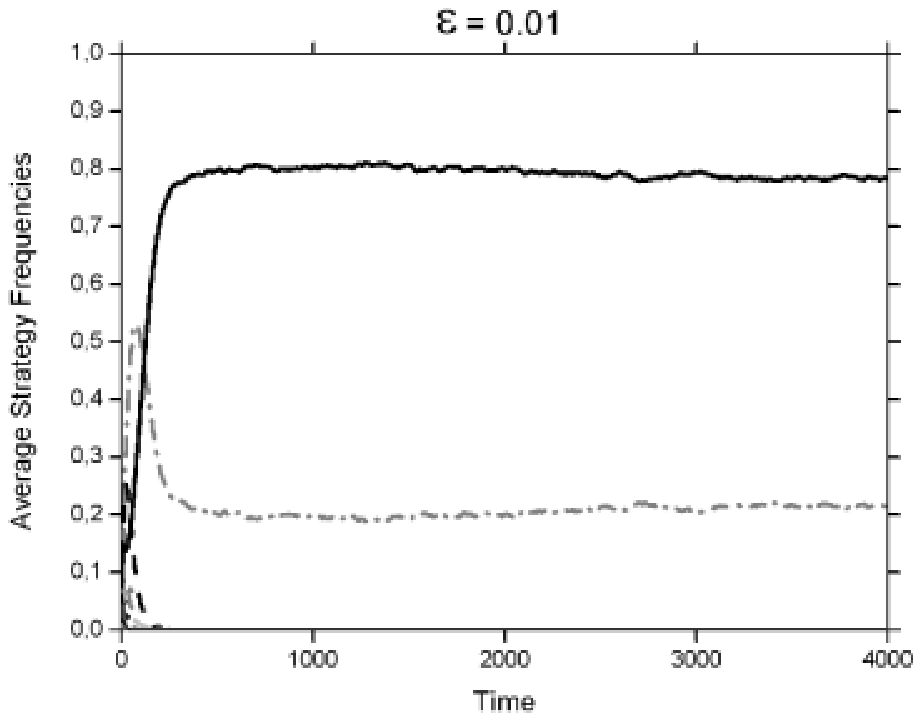} &
\includegraphics[width=0.5\textwidth,height=!]{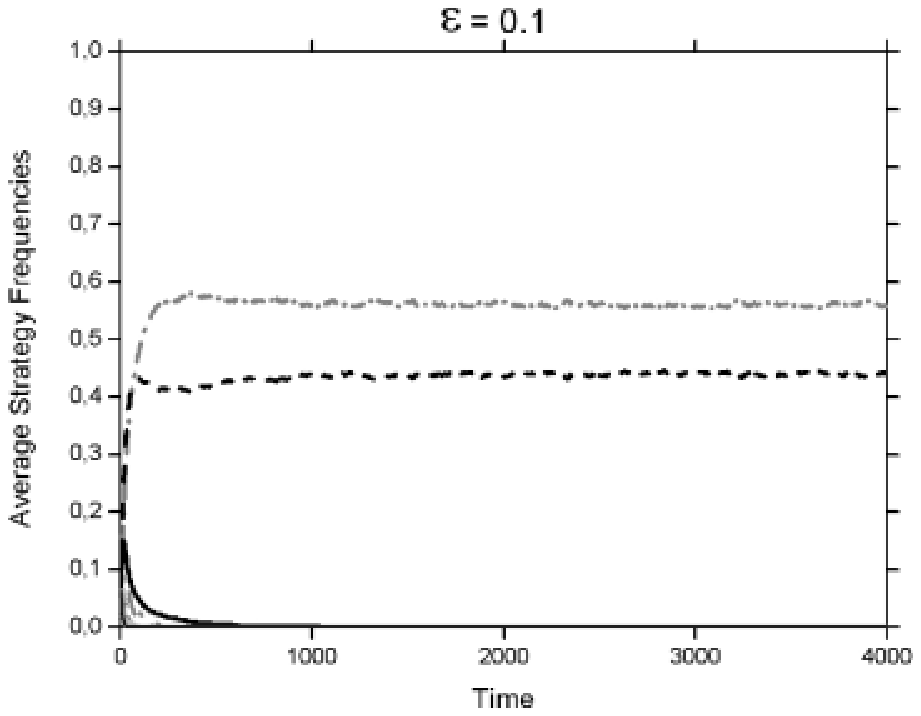}
\end{tabular}
\caption{Frequencies for the 16 competing BMS vs the number of time steps
for Deterministic Hero Game with $\tilde{S}=1.333$,
$\tilde{R}=0.5$ and $q=8$.
Color and line style codes are the same than in Figure \ref{fig:z4det4ep}.}
\label{fig:z8hero}
\end{figure}
\end{center}

Fig. \ref {fig:z8hero} shows the results the Hero deterministic game with 
$q=8$.
For the case without noise ($\epsilon=0$), we have two main strategies 
:[0,0,1,1] and [1,0,1,1]. For intermediate amounts of noise 
$\epsilon << 0.1$, the strategy [1,0,1,1] takes over.
Finally, for large amounts 
of noise ($\epsilon\simeq0.1$) we have a new winner strategy: 
[1,0,0,1] (PAVLOV). 

For the Stag Hunt game (not plotted here), the dominant strategy is 
always [1,1,1,1] (ALLWAYS C), a result that can be explained because 
for this game playing C pays back a lot.

\section{Discussion}

We developed a simple model to study evolutionary strategies in spatial $2 
\times 2$ games that provides more robust results than those from 
more complex previous models \cite{bkd99}.
We found few dominant strategies that appear repeatedly for 
several different $2 \times 2$ games, 
and not only for the Prisoner's Dilemma. Comparing figures 
(\ref{fig:z4det4ep})-(\ref{fig:z8hero}), 
we notice that 3 strategies -mainly [1,0,1,1] and less often 
[0,0,1,0] and [0,0,1,1]- dominate for the
different games, update rules and noise levels.
If we look for these strategies at Table \ref{scores}, we observe that 
none of them are "winner" strategies in the non spatial games. 
That is, none of them get the highest average payoff but just a 
mediocre one.
So territoriality seems to have a relevant effect on the evolution of 
strategies. 
Moreover, the departure from the non spatial tournament becomes larger
as the neighborhood size grows.

Another important conclusion is that for a large enough level of noise 
the diversity disappears and one ends with just one universal strategy 
( mainly [1,0,1,1] )
or at most two dominant but non coexisiting strategies.

The strategy [1,0,1,1] is particularly interesting because it is like a 
"crossover" between  PAVLOV [1,0,0,1] and TFT [1,0,1,0], which are the 2 
main strategies that humans use when are engaged in social dilemma game
experiments \cite{wm96},\cite{mw98}.
We baptized this strategy as the 
"Non-Tempted Pavlov".

From the different variations of our model, we found that the evolution of 
more cooperative strategies (more conditional probabilities $p_X$ equal to 1)
is favored when:
 
\begin{itemize}

\item The size of the neighborhood is increased 
($q=8$ lead to dominant strategies with
much no-null conditional probabilities than $q=4$).

\item The update rule version for $c(x,y;t)$ after the agent at ($x,y$) 
played with its $q$ neighbors is deterministic. 

\item The amount of noise measured by $\epsilon$ increases. 

\end{itemize}

Some issues that deserve further study is to use time 
integrated utilities instead of the instantaneous 
utilities used here, also to analyze spatial patterns 
(for instance, size and form of cooperative clusters and of 
winning strategy clusters).

\section{Acknowledgements}

We are greatful to Daniel Ariosa and Michael Doebeli for useful comments.

\end{document}